\documentclass[12pt, preprint]{aastex}
\def\deg{\ifmmode^\circ\else$^\circ$\fi}

\def\mic{~$\mu$m}

\def\mic{$\mu${\rm m}}

\def\arcs{\ifmmode {''}\else $''$\fi}
\def\arcm{\ifmmode {'}\else $'$\fi}
\def\parcs{\sa=.07em \sb=.03em
     \ifmmode $\rlap{.}$^{\scriptscriptstyle\prime\kern -\sb\prime}$\kern -\sa$
     \else \rlap{.}$^{\scriptscriptstyle\prime\kern -\sb\prime}$\kern -\sa\fi}
\def\parcm{\sa=.08em \sb=.03em
     \ifmmode $\rlap{.}\kern\sa$^{\scriptscriptstyle\prime}$\kern-\sb$
     \else \rlap{.}\kern\sa$^{\scriptscriptstyle\prime}$\kern-\sb\fi}

\def\Lsun{L$_{\odot}$}

\def\spose#1{\hbox to 0pt{#1\hss}}
\def\simlt{\mathrel{\spose{\lower 3pt\hbox{$\mathchar"218$}}
     \raise 2.0pt\hbox{$\mathchar"13C$}}}
\def\simgt{\mathrel{\spose{\lower 3pt\hbox{$\mathchar"218$}}
     \raise 2.0pt\hbox{$\mathchar"13E$}}}
\def\lsim{\rlap{$<$}{\lower 1.0ex\hbox{$\sim$}}}
\def\gsim{\rlap{$>$}{\lower 1.0ex\hbox{$\sim$}}}

\begin{document}

\title{Silicate Emission in the {\it Spitzer}$^1$\ IRS$^2$\ spectrum of FSC 10214+4724}

\altaffiltext{1}{based on observations obtained with the {\it Spitzer Space Telescope}, which is 
operated by JPL, California Institute of Technology for the National Aeronautics and Space Administration}

\altaffiltext{2}{The IRS is a collaborative venture between Cornell
 University and Ball Aerospace Corporation that was funded by NASA
   through JPL.}

\author{H. I. Teplitz\altaffilmark{3}, 
L. Armus\altaffilmark{3},
B.T. Soifer\altaffilmark{3},
V. Charmandaris\altaffilmark{4,5,6},
J. A. Marshall\altaffilmark{4},
H. Spoon\altaffilmark{4},
C. Lawrence\altaffilmark{7},
L. Hao\altaffilmark{4},
S. Higdon\altaffilmark{4},
Y. Wu\altaffilmark{4},
M. Lacy\altaffilmark{3},
P. R. Eisenhardt\altaffilmark{7},
T. Herter\altaffilmark{4},
J.R. Houck\altaffilmark{4}
}

\altaffiltext{3}{Spitzer Science Center, MS 220-6, Caltech, Pasadena, CA 91125.  hit@ipac.caltech.edu}
\altaffiltext{4}{Astronomy Department, Cornell University, Ithaca, NY  14853}
\altaffiltext{5}{Chercheur Associ\'e, Observatoire de Paris, F-75014, Paris, France}
\altaffiltext{6}{University of Crete, Dept. of Physics, GR-71003 Heraklion, Greece}
\altaffiltext{7}{Jet Propulsion Laboratory, California Institute of Technology, 4800 Oak Grove Drive, Pasadena, CA 91109}

\begin{abstract}
  
  We present the first MIR spectrum of the $z=2.2856$ ultraluminous,
  infrared galaxy FSC 10214+4724, obtained with the Infrared
  Spectrograph onboard the {\it Spitzer Space Telescope}.  The
  spectrum spans a rest wavelength range of $2.3-11.5\mu$m, covering a
  number of key diagnostic emission and absorption features.  The most
  prominent feature in the IRS spectrum is the silicate emission at
  rest-frame $\sim 10$\ \mic.  We also detect an unresolved emission
  line at a rest wavelength of $7.65\mu$m which we identify with
  [NeVI], and a slightly resolved feature at 5.6 \mic\ identified as a
  blend of [Mg VII] and [Mg V].  There are no strong PAH emission
  features in the FSC 10214+4724 spectrum.  We place a limit of
  $0.1\mu$m on the equivalent width of $6.2\mu$m PAH emission but see
  no evidence of a corresponding 7.7 \mic\ feature.  Semi-empirical
  fits to the spectral energy distribution suggest $\sim 45$\% of the
  bolometric luminosity arises from cold ($\sim 50$\ K) dust, half
  arises from warm (190 K) dust, and the remainder, $\sim 5\%$
  originates from hot ($\sim 640$\ K) dust.  The hot dust is required
  to fit the blue end of the steep MIR spectrum.  The combination of a
  red continuum, strong silicate emission, little or no PAH emission,
  and no silicate absorption, makes FSC 10214+4724 unlike most other
  ULIRGs or AGN observed thus far with IRS.  These apparently
  contradictory properties may be explained by an AGN which is highly
  magnified by the lens, masking a (dominant) overlying starburst with
  unusually weak PAH emission.

 \end{abstract}


\section{Introduction}

FSC 10214+4724, at a redshift of $z=2.2856$\ \citep{Rowan-Robinson
  1991}, was initially thought to be the most luminous object in the
Universe, but was later revealed to be gravitationally lensed by a
foreground galaxy \citep{Broadhurst 1995, Graham 1995, Serjeant 1995,
  Eisenhardt 1996}.  The lensing model of \cite{Eisenhardt 1996}\ 
suggests that the central (optical/UV) source is magnified by a factor
of $\sim 100$, and that the lensing arc is an image of the central
$\sim 0^{\prime \prime}.005$\ (40 pc) of the source at B-band rest
wavelength.  They further conclude that the bolometric luminosity of
the source is produced over a larger region (240 pc), implying a
bolometric magnification factor of 30 and an intrinsic luminosity of
$\sim 2 \times 10^{13}$\ \Lsun\ , placing it in the class of
ultraluminous infrared galaxies (ULIRGs).  The large magnification
makes FSC 10214+4724 a unique subject for the study of high redshift
ULIRGs, offering greater effective sensitivity and spatial resolution
than possible in observations of unlensed sources.

A central question in understanding FSC 10214+4724 is the relative
contribution of starburst and AGN components to its luminosity.  The
rest-frame UV-optical spectrum appears to be similar to that of
Seyfert 2 galaxies \citep{Elston 1994}.  Strong UV polarization shows
that much of the UV continuum results from scattered light, and broad
lines in the polarization spectrum identify the presence of an AGN
\citep{Lawrence 1993, Goodrich 1996}. The bulk of the luminosity,
however, is emitted in the infrared \citep[as much as
99\%,][]{Rowan-Robinson 1991}.  CO observations point to a large
reservoir of molecular gas \citep{Solomon 1992, Scoville 1995}.
Sub-mm and millimeter detections \citep{Downes 1992, Rowan-Robinson
  1993,Benford 1999} suggest substantial emission from cold dust,
usually associated with extended star formation.  Recent {\it
  Chandra}\ observations show the object to have weak X-ray emission,
consistent with vigorous star formation or a Compton-thick AGN \citep{Alexander 2005}.  Lens
models suggest differential amplification, so that the central
(AGN-dominated) region is more highly magnified than the surrounding
starburst \citep[e.g.,][]{Eisenhardt 1996, Lacy 1998}.

The sensitivity and large wavelength coverage of the {\it Spitzer}\ 
Infrared Spectrograph \citep[IRS][]{Houck 2004}\ makes it possible to
explore the dust emission and absorption features in the rest-frame
mid-infrared spectra of dusty galaxies at low and high redshift, and
thus identify the power sources which may be hidden in the UV and
optical.  In this paper, we present the first MIR spectrum of FSC
10214+4724.  We describe the observations and data reduction in
Section 2, present the spectrum and a dust emission model fit to the
spectral energy distribution in Section 3, and discuss the
implications in Section 4.  Throughout, we assume a
$\Lambda$-dominated flat universe, with $H_0=70$\ km s$^{-1}$\ 
Mpc$^{-1}$, $\Omega_{\Lambda}=0.7, \Omega_{m}=0.3$.

\section{Observations and Data Reduction}

Spectra of FSC 10214+4724 were obtained with the IRS on 19 April 2004.
The data were taken in the first order of the low resolution, short
wavelength module (SL-1; 7.5-14.2 \mic) and in both orders of the low
resolution, long wavelength module (LL-1 and LL-2, 14.2-21.8 and
20.6-38 \mic, respectively).  The spectral resolution varies from 60
to 120 across each order.  Individual ramp durations were 60 seconds
in SL-1 and 120 in LL-1 and LL-2.  Spectra were taken in the standard
``staring'' mode, with four exposures at each of two positions,
separated by one third of the slit length.  A total of 8 individual
spectra were taken in each sub-slit, for total on-source integration
times of 480, 960, and 960 seconds.

Spectra were reduced using the S11 pipeline at the {\it Spitzer}\ 
Science Center, which includes ramp fitting, dark sky subtraction,
droop correction, linearity correction, and wavelength calibration.
One-dimensional spectra were extracted from the un-flatfielded
two-dimensional spectra using the SMART data reduction package
\citep{Higdon 2004}.  The extractions are then flux calibrated with
the IRS spectrum of a star ($\alpha$\ Lac), extracted in an identical
manner using SMART.  The data have been sky-subtracted by
differencing the two nod positions along the slit, before spectral
extraction.  As a final step, we have scaled the SL-1 and LL-2 1D
spectra by 3\% and 8\%, respectively, to match the LL-1 spectrum in
the overlap region and produce a single low-resolution spectrum from
$7.5-38\mu$m (observed frame).

Mid-IR photometry of FSC10214+4724 at 3.6, 4.5, 5.8 and 8.0 $\mu$m was
obtained on 20 May 2004 using the Infrared Array Camera
\citep[IRAC;][]{Fazio 2004}. The galaxy was observed in full array
mode with a cyclic 5 point dither pattern resulting in a total on
source integration time of 1 minute per filter. The source was
unresolved and its Full-Width at Half Maximum (FWHM) varied between
1.5\arcsec and 2\arcsec. We performed photometry on the final mosaics
produced by the SSC pipeline (S11.0.2) using an aperture of 3.6 arcseconds
radius following the method described in the IRAC data Handbook.
The resulting flux densities are accurate to $<$5\%.  No attempt was made to 
correct the IRAC photometry for the contribution of the lensing
galaxy which is unresolved from the source.

\section{Results}

\begin{figure}[t!]
\plotone{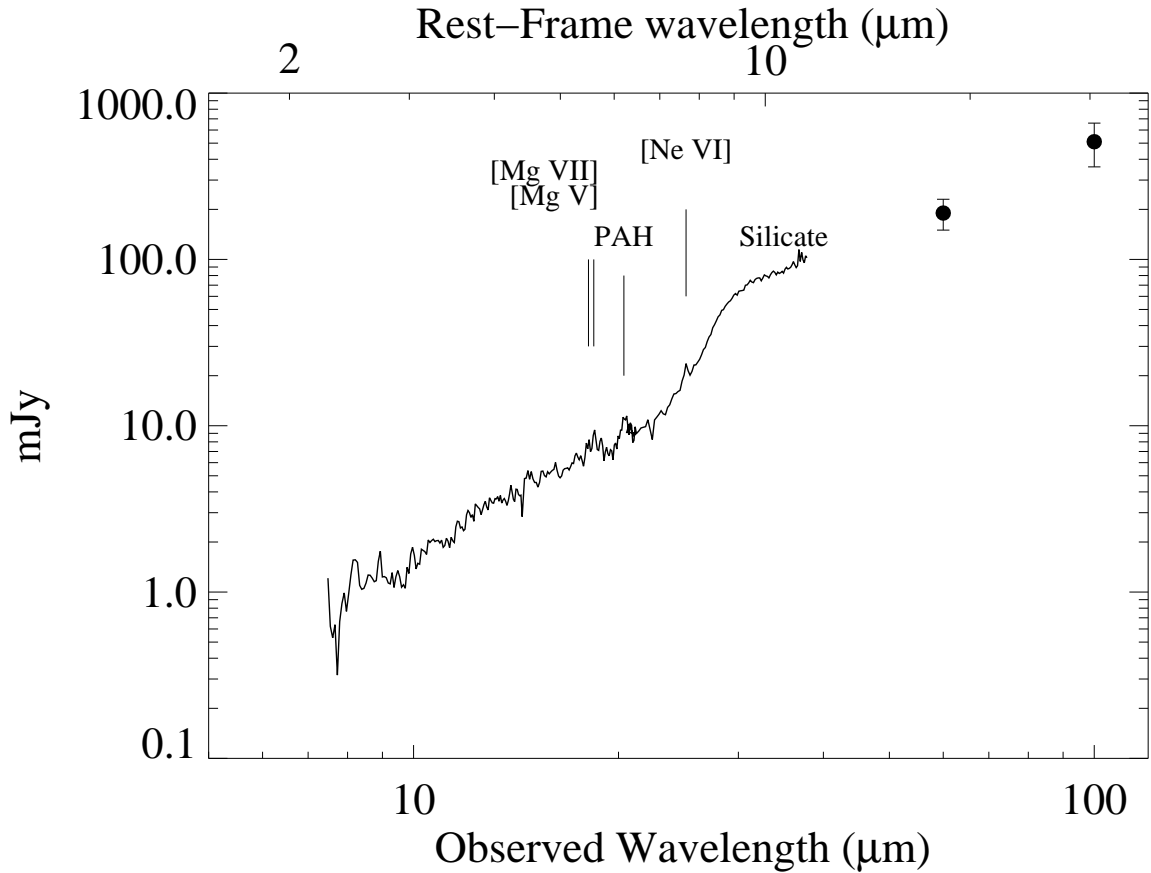}
\caption{\label{fig: spectrum} The IRS spectrum and IRAS photometry of FSC 10214+4724.  The extraction from
  the three modules, SL-1, LL-2 and LL-1, have been ``stitched''
  together, cropping the noisier overlap wavelength regions.  The
  filled circles show the IRAS photometry.  Identified emission
  features are labeled.  }
\end{figure}

We plot the IRS spectrum of FSC 10214+4724 in Figure \ref{fig: spectrum}.  
The most prominent feature is a steep rise at the red end ($\sim
8-10$\ \mic\ rest wavelength).  Extending the spectral energy
distribution (SED) with the IRAS 60 and 100 \mic\ photometry
\citep{Moshir 1990} makes clear that there is not simply a steeply
rising continuum, but rather is a broad emission feature on top of a
somewhat less steep continuum.  We identify this feature as silicate
emission, centered at $\sim 10$\ \mic.  It is difficult to measure an
accurate equivalent width for the feature, because the IRS wavelength
coverage does not extend far enough to the red.
 
Three weak emission lines are also present.  We identify the narrow
emission line at rest wavelength 7.65 \mic\ as the [Ne VI] fine
structure line.  This line has a rest-frame equivalent width (EW) of
0.02 \mic, roughly comparable to that observed in low redshift
Seyferts \citep{Sturm 1999, Lutz 2000}.  The emission feature observed
at $\sim 5.5$\ \mic\ in the rest-frame appears to be marginally
resolved.  We identify it as a blend of [Mg VII] and [Mg V] at 5.503
and 5.610 \mic, respectively.  These lines have been seen in nearby
AGN \citep{Sturm 2002}\ and have ionization potentials of about 186
and 105 eV, similar to the 158 eV ionization potential of [Ne VI].
While there is also a rotational transition of H$_{2}$ at $5.51\mu$m
(the S(7) line) which is often quite strong in ULIRGs, this feature is
always accompanied by much stronger emission from the other, lower
transition, H$_{2}$ rotational lines, which are not present.  The
rest-frame equivalent width of the [Mg VII] and [Mg V] lines combined
is $\sim 0.03\pm 0.01$\ \mic.  While uncertain, this EW appears to be
a factor of 3-5 higher than in local AGN \citep{Sturm 2002}.  The EW
of the Ne and Mg lines is a factor of a few higher than the strongest
of the narrow, high-excitation emission lines (C IV, He II, [Ne IV])
seen in the rest-frame UV spectrum of FSC 10214+4724
\citep{Rowan-Robinson 1991, Goodrich 1996}.

There is a marginal detection of broad emission at 6.2 \mic\ 
rest-frame, corresponding to the wavelength of polycyclic aromatic
hydrocarbon (PAH) emission.  Although the feature is clearly visible
in Figure \ref{fig: spectrum}, examination of the individual
extractions shows that it is more prominent in LL-2 than in LL-1.  The
observed wavelength (20.3 \mic) places the feature near the noisy blue
end of LL-1, but at a wavelength that is usually regarded as reliable.
We take the measurement, $EW \sim 0.1$\ \mic\ in the rest-frame,
as an upper limit.  We also note that the corresponding 7.7 and 8.6
\mic\ PAH emission features are not seen, despite being redshifted
into a clean part of the spectrum and the fact that they are usually
1.5--2 times stronger than the 6.2 \mic\ feature in
starburst-dominated low redshift ULIRGs.  The low ratio of 7.7 to 6.2
\mic\ PAH equivalent width, while highly unusual, is not impossible
under certain conditions \citep{Kessler-Silacci 2005}.  The 11.3 \mic\ 
PAH feature might also be expected, but that line falls near the red end
of LL-1 where the noise precludes a meaningful limit.  Nonetheless,
the lack of other PAH lines indicates that the 6.2 \mic\ feature
should be treated with caution.

\section{Discussion}

The shape of the SED of FSC 10214+4724 is dominated by
dust at a variety of temperatures.  Substantial
cold dust must be present, given the strong emission at long
wavelengths \citep[$> 40$\ \mic, rest-frame; e.g., ][]{Rowan-Robinson 1993}.
At the same time, warmer dust will be required to explain the
5-10 \mic\ continuum.  Dust warmer than 100 K can produce 
the silicate emission \citep[e.g.,][]{Li 2001}, but a hot dust
component (several hundred K) is required to explain the $\sim 5$\ 
\mic\ continuum.  The rest-frame near-infrared (NIR) will have a contribution 
from both star light and hot dust.

\begin{figure}[t!]
\plotone{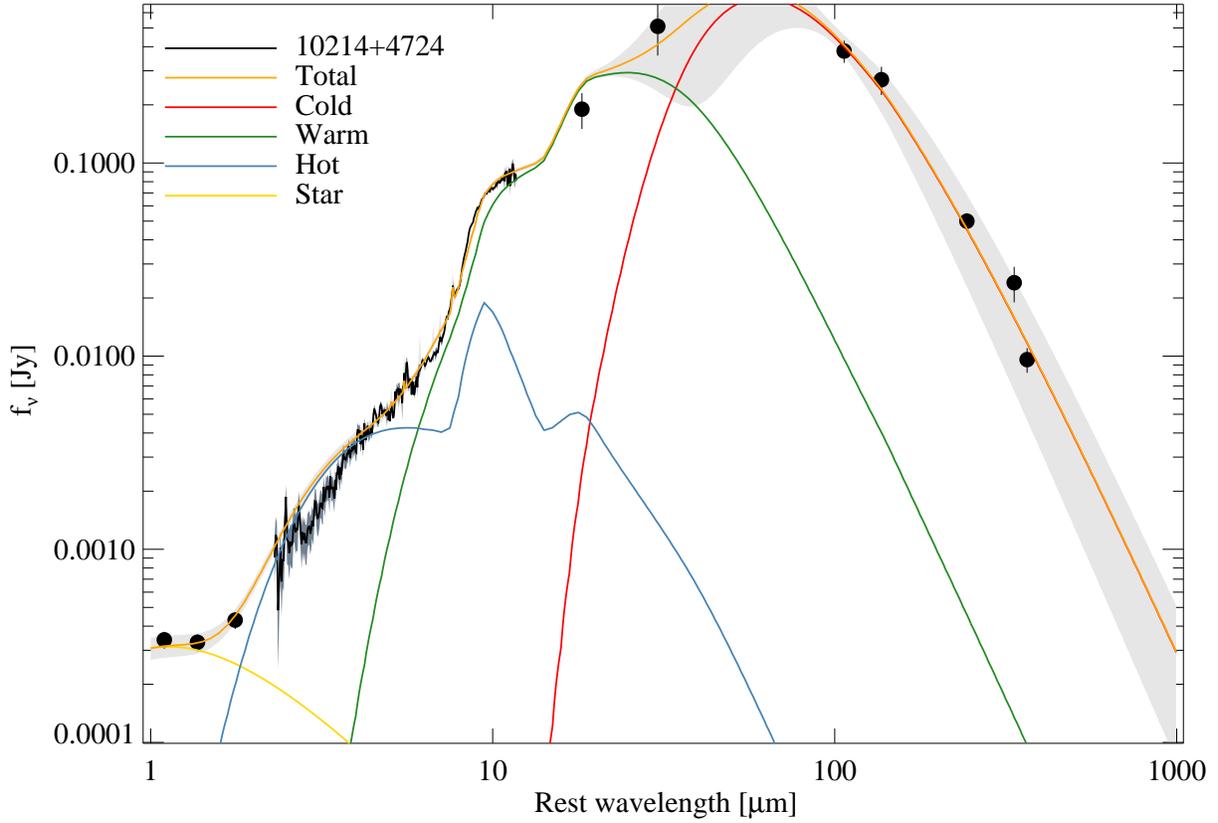}
\caption{ \label{fig: SED fit} The semi-empirical dust model fit to the SED of FSC 10214+4724.  The
  IRS mid-IR spectrum was extended with other data as described in the
  text.  The components of the model (cold, warm, and hot dust, and
  stellar 3500K black body) are indicated in color.  The shaded grey
  region indicates the $1\sigma$\ uncertainty in the fit.  }
\end{figure}

We have fit the SED with a
multi-component model which includes three graphite and silicate dust
grain components and a 3500K blackbody stellar component \citep{Marshall 2006}.
The three dust components in the model are not meant to represent
three distinct physical structures, but rather they are indicative of
the average temperature ranges within the source.  The model was fit
to the SED from observed frame NIR to millimeter wavelengths (see
Figure \ref{fig: SED fit}.  NIR data included the photometry of
\cite{Soifer 1991}\ and the IRAC data described in Section 2.  Longer
wavelength data included IRAS photometry at 60 and 100 \mic, the 350
\mic\ detection of \cite{Benford 1999}, and the sub-mm and mm data of
\cite{Rowan-Robinson 1993}\ and \cite{Downes 1992}.

A minimum of three dust components are required to fit the FSC
10214+4724 SED, one at $\sim 50$\ K (cold), one at $\sim 190$\ K
(warm), and one at $\sim 640$\ K (hot). Their relative contributions
to the bolometric luminosity are given in Table \ref{tab: SED fit} .   
Note that hot dust contribution is an upper limit, because it
is dominated by the IRAC data and no correction has been made for
contamination by the foreground lensing galaxy.
The stellar emission is a negligible contribution to the bolometric
luminosity, but is needed to fit the shortest wavelength data.
component.  The cold dust component ($51 \pm 6$\ K) is in good
agreement with the estimate of \cite{Benford 1999}, 55 K.  Each
component contains a distribution of grains with different equilibrium
temperatures depending on their size and composition.  Additionally,
each component contains emission from dust at different radial
distances, and therefore temperatures, from the illuminating source.
We define the characteristic temperature of a component to be the
temperature of the most luminous grain size at the distance from the
source contributing the majority of the luminosity.  This luminosity
dominating distance corresponds to a $\tau(UV) \sim 0.5$, where
approximately half of the UV-source photons have been absorbed.  Each
component therefore contains dust above and below the characteristic
temperature.  With this definition, the characteristic temperature
roughly corresponds to the expected peak in the dust modified
(grey-body) Planck function.

The observed SED of FSC 10214+4724 appears consistent with low
redshift, AGN-dominated ULIRGs.  Such sources can have cold dust
components which account for up to 40\%\ of the bolometric luminosity,
due to both AGN-heated cold dust far from the nucleus and the presence
of a small underlying starburst \citep{Armus 2004, Armus 2006}.
\cite{Rowan-Robinson 2000} also estimates that the AGN contributes
most of the luminosity of FSC 10214+4724.  In addition, FSC 10214+4724
shows silicate emission, similar to other AGN observed with the IRS,
and a hot (graphite) dust continuum.  However, the hot dust
contribution is quite small compared to many AGN-dominated ULIRGs or
QSOs.  The only obvious fine-structure lines are [Ne VI], [Mg VII] and
[Mg V], high-ionization species not observed in starburst galaxies.
There is little if any PAH emission.  The upper limit to the $6.2\mu$m
EW is approximately a factor of $5-10$ lower than most pure starburst
galaxies \citep{Brandl 2005}\ and ULIRGs dominated by star formation
\citep{Armus 2004, Armus 2006}.

However, differential magnification is likely to be enhancing the central AGN,
reducing the PAH EW and making the silicate emission more obvious in
the IRS spectrum.  \cite{Eisenhardt 1996}\ estimate a magnification
factor of $\sim 100$\ for the central 40 pc, but only a factor of
$\sim 30$\ out to at least 240 pc.  The intrinsic contribution of the
starburst to the bolometric luminosity could be larger than that
inferred by our current model.  We assume that the warm and hot dust
heated by the AGN are magnified by an additional factor of 3.3, but
that both the cold dust heated by the AGN and any dust heated by
starburst are not.  The ratio of cold to warm dust in local AGN- and
starburst-dominated ULIRGs is approximately 2:3 and 3:1, respectively
(Armus et al. 2006, in preparation).  Taking these assumptions
together, we estimate the intrinsic AGN contribution to the bolometric
luminosity of FSC 10214+4724 to be $\sim 35$\% after correction for
the differential magnification.  If the differential magnification
factor is correct, this AGN contribution is probably an upper limit 
because AGN-dominated ULIRGs often have measurable starbursts.

If the starburst contributes 65\%\ of the intrinsic luminosity, it is
surprising that little or no PAH emission is seen.  Taking our limit
of 0.1 \mic\ $>EW_{\mbox{rest}}$\ and a differential magnification
factor of 3.5 would place the limit of the EW for PAH emission within
the range of other star-forming ULIRGs\citep[Armus et al.  2005, in
prep;][]{Brandl 2005}.  However, the uncertainty in our PAH measurement leaves
this possibility unconfirmed; a lower signal to noise ratio would have
made our limit higher.  Furthermore, the lack of evidence for
comparably strong PAH emission at 7.7 \mic\ makes it likely that the
EW at 6.2 \mic\ is overestimated.  Nonetheless, the limit demonstrates
that substantial star formation can be present in FSC 10214+4724 and
not visible in the (differentially magnified) IRS spectrum.

\begin{figure}[]
\plotone{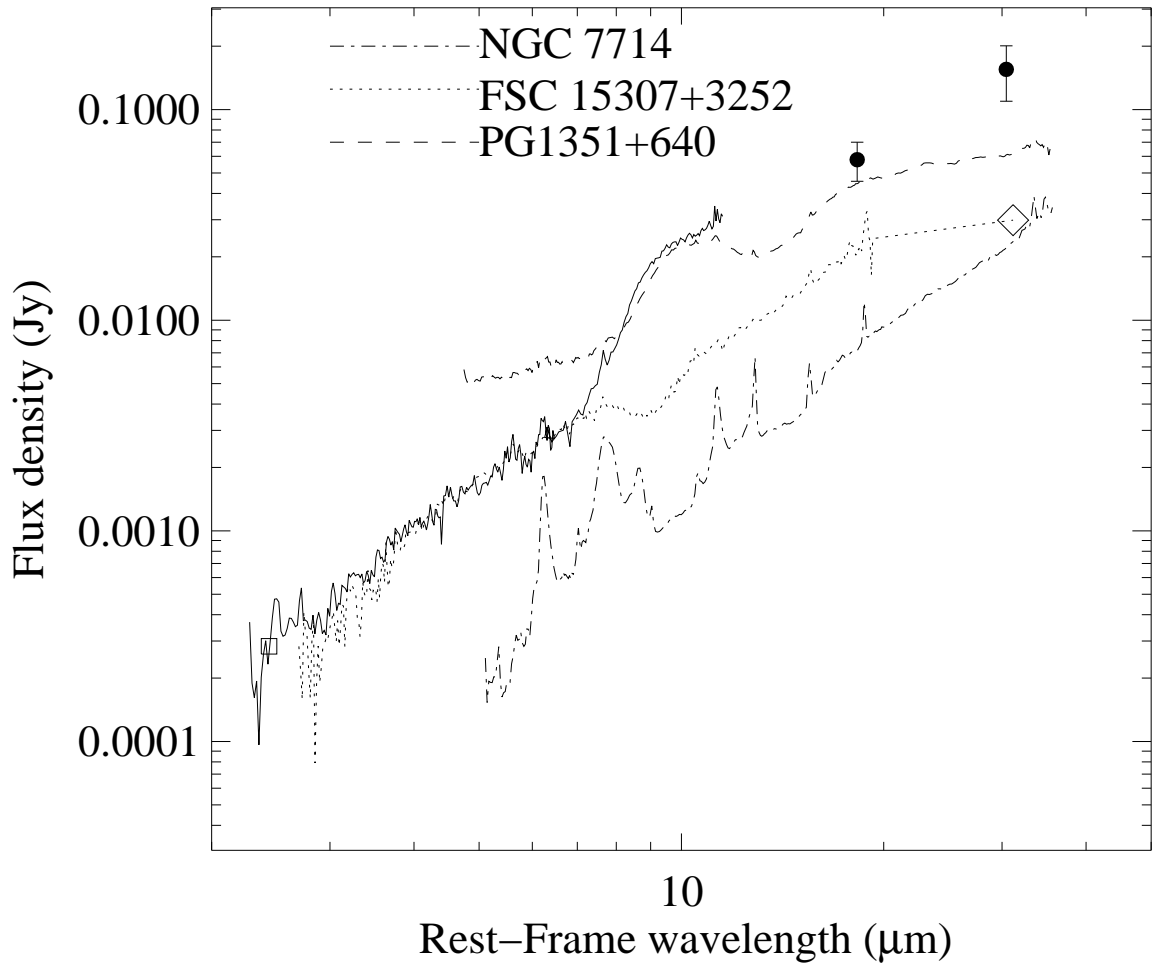}
\caption{\label{fig: SED compare} The SED of FSC 10214+4724 compared to
  other dusty galaxies observed with the IRS.  We show the MIR
  spectrum of FSC 10214+4724 (solid line) and IRAS photometry (solid
  circles with error bars).  For comparison, we overplot: the IRS
  spectrum of PG 1351+640 \citep[dashed line;][]{Hao 2005}, normalized
  to the prominent silicate emission; the IRS spectrum of the ULIRG
  FSC 15307+3252 (dotted line; Charmandaris 2005, in prep.) and its
  IRAS photometry ( connected by dotted line, data point shown as open
  diamond) normalized at 5-7 \mic; and the spectrum of NGC 7714
  \citep[dot-dashed line;][]{Brandl 2004}, normalized at 5-7 \mic\ and
  then offset by a factor of 5 for clarity.  Note that the
  differential magnification causes the observed spectrum of FSC
  10214+4724 to appear bluer than it would intrinsically, so the
  comparison in this figure should be taken as a lower limit.}
\end{figure}

In figure \ref{fig: SED compare}, we compare the rest frame
MIR spectrum of FSC 10214+4724 to the IRS spectra of three
other sources.  The intrinsic spectrum of FSC 10214+4724 is redder
than the observed spectrum, given the differential magnification, so
the differences seen in the figure would be greater if corrected for
lensing.  The comparison sources are an AGN-dominated ULIRG (FSC
15307+3252), a silicate-emitting QSO (PG 1351+640) and a pure
starburst galaxy (NGC 7714).  PG 1351+640 has similar silicate
emission, but has a much stronger hot dust continuum at $\sim 5$\ 
\mic\ \citep{Hao 2005}.  In fact, the cold dust emission in FSC
10214+4724 gives it a much steeper mid-to-far infrared slope than most 
AGN with silicate emission.  The 30 \mic\ to 6
\mic\ flux density ratio of 46 in FSC 10214+4724, is higher than the
reddest of the quasars in \cite{Hao 2005}\ or \cite{Siebenmorgen
  2005}, with a ratio of 11, and the reddest AGN in \cite{Weedman
  2005}, NGC 1275, which has a ratio of 30.  The ULIRG FSC 15307+325
has a similar MIR slope to FSC 10214+4724 in the 2-10 \mic\ region,
but has silicate absorption rather than emission, and a much weaker
cold dust continuum at longer wavelengths.  The continuum of NGC7714
is very red with a 30 \mic\ to 6 \mic\ flux density ratio of 120, and
a strong stellar contribution at the short wavelengths.

In many AGN-dominated ULIRGs, the MIR spectrum is even bluer, with a
strong continuum in the $\sim 5$\ \mic\ region \citep[e.g.,
][]{Laurent 2000}.  ULIRGs with steeper mid-infrared continua tend
also to have silicate absorption, not emission.  We see no evidence in
the spectrum of FSC 10214+4724 for an underlying silicate absorption feature
at 9.7 \mic.
Furthermore, the profile of the silicate emission is consistent with
that seen in most other AGN observed with the IRS \citep{Hao 2005,
  Siebenmorgen 2005, Weedman 2005}; although see \cite{Sturm 2005}\ 
for a counter example and discussion of the factors influencing the
profile shape.  While some absorption may be filled in with emission,
the absorption and emission profiles are not necessarily identical.  A
coincidence of opacity, temperature, and grain mixture in the emitting
and absorbing regions would be required.

The AGN contribution to the bolometric luminosity in FSC 10214+4724 is
apparently substantial but does not dominate.  Nonetheless, the PAH
emission expected for a starburst is weak or absent.  An unusual dust
geometry would be required for an AGN alone to explain the SED, given
the amount of cold dust.  The other available evidence points to a
standard AGN configuration.  The shape of the spectrum between
$2-12\mu$m is qualitatively similar to dusty torus models \citep{Pier
  and Krolik 1992}\ with inner radius to height ratios of $a/h\sim
0.3$, and opening angles of $\sim 50$\deg -- consistent with a
high-luminosity equivalent of Seyfert 2 galaxy, like NGC 1068. The
similarlity to NGC 1068 hass previously been
noted\citep[e.g.,][]{Barvainis 1995}.  The presence of silicate
emission in the IRS spectrum rules out models where the torus is seen
edge-on, or where the AGN is completely obscured by foreground (cold)
dust.

\acknowledgements

This work is based in part on observations made with the {\it Spitzer
  Space Telescope}, which is operated by the Jet Propulsion
Laboratory, California Institute of Technology under NASA contract
1407. Support for this work was provided by NASA through an award
issued by JPL/Caltech.

\clearpage

\begin{deluxetable}{lll}
\tablecaption{SED fit parameters \label{tab: SED fit}}
\tablehead{
\colhead{Dust Component} &
\colhead{Temperature (K) } &
\colhead{$L/L_{tot}$\ (\%)} 
}

\startdata

Hot  & $638\pm20$ & 5.6 \\
Warm & $191\pm 2$ & 51.3 \\
Cold & $51\pm 6$  & 43.3 

\enddata

\end{deluxetable}

\clearpage

\clearpage

\clearpage

\end{document}